\documentclass[pra,aps,superscriptaddress,twocolumn,noshowpacs]{revtex4}%

\usepackage{graphicx}
\usepackage[normal]{subfigure}
\usepackage{caption}
\usepackage{latexsym}
\usepackage{amsmath}
\usepackage{amssymb}
\usepackage{amsfonts}
\usepackage{color}
\usepackage{dsfont}
\usepackage{caption}

\begin{document}

\newcommand{\x}{\hat{x}_1}
\newcommand{\xx}{\hat{x}_2}
\newcommand{\p}{\hat{p}_1}
\newcommand{\pp}{\hat{p}_2}
\newcommand{\ket}[1]{|#1\rangle}
\newcommand{\bra}[1]{\langle#1|}
\newcommand{\moy}[1]{\langle#1\rangle}
\newcommand{\tr}{\mathrm{Tr}}


\title{Detection of non-Gaussian entangled states \\ with an improved continuous-variable separability criterion}

\author{Anaelle Hertz}
\email{ahertz@ulb.ac.be}
\affiliation{Centre for Quantum Information and Communication, \'Ecole polytechnique de Bruxelles, CP 165, Universit\'e libre de Bruxelles, 1050 Brussels, Belgium}

\author{Evgueni Karpov}
\affiliation{Centre for Quantum Information and Communication, \'Ecole polytechnique de Bruxelles, CP 165, Universit\'e libre de Bruxelles, 1050 Brussels, Belgium}

\author{Aikaterini Mandilara}
\affiliation{Department of Physics, School of Science and Technology, Nazarbayev University, 53, Kabanbay
Batyr Avenue, Astana, 010000, Republic of Kazakhstan}

\author{Nicolas J. Cerf}
\affiliation{Centre for Quantum Information and Communication, \'Ecole polytechnique de Bruxelles, CP 165, Universit\'e libre de Bruxelles, 1050 Brussels, Belgium}


\begin{abstract}

  Currently available separability criteria for continuous-variable states are generally based on the covariance matrix of quadrature operators. The well-known separability criterion of Duan {\it et al.} [Phys. Rev. Lett. {\bf 84}, 2722 (2000)] and Simon [Phys. Rev. Lett. {\bf 84}, 2726 (2000)] , for example, gives a necessary and sufficient condition for a two-mode Gaussian state to be separable, but leaves many entangled non-Gaussian states undetected. Here, we introduce an improvement of this criterion that enables a stronger entanglement detection. The improved condition is based on the knowledge of an additional parameter, namely the degree of Gaussianity, and exploits a connection with Gaussianity-bounded uncertainty relations [Phys. Rev. A {\bf 86}, 030102 (2012)]. We exhibit families of non-Gaussian entangled states whose entanglement remains undetected by the Duan-Simon criterion.

\end{abstract}

\maketitle

\section{Introduction}

Quantum entanglement is nowadays considered a central resource in the field of quantum information and computation  \cite{horo}. It is  therefore crucial to be able to determine whether a quantum state is separable or entangled, which is provably a hard decision problem when it comes to mixed states. In the context of continuous-variable systems, such as bosonic modes or collective atomic spins (in the limit of large ensembles), a necessary criterion for the separability of any two-mode state has been derived by Duan {\it et al.} \cite{duan}  and Simon \cite{simon}, which even turns into a necessary and sufficient criterion in the special case of Gaussian states. This criterion results from translating to continuous-variable (infinite-dimensional) systems the positive partial transpose (PPT) condition, which had been established for finite-dimensional discrete systems \cite{peres, horo2}. 
 Following the notation of Duan {\it et al.} \cite{duan}, the criterion expresses that if a two-mode state is separable, then its so-called Einstein-Podolsky-Rosen (EPR) variance complies with the  inequality
 \begin{equation}
\Delta\equiv\frac{1}{2}\Big( \langle(\Delta\hat{u})^2\rangle+ \langle(\Delta\hat{v})^2\rangle\Big)\geq\frac{1}{2}\Big(\alpha^2+\frac{1}{\alpha^2}\Big),
\label{delta}
 \end{equation}
for any real (nonzero) $\alpha$, where the operators
 \begin{equation}
 \hat{u}=|\alpha|\x+\frac{1}{\alpha}\xx,\qquad\quad\hat{v}=|\alpha|\p-\frac{1}{\alpha}\pp,
 \end{equation}
are functions of the quadratures components $\hat{x}$ and $\hat{p}$ of modes 1 and 2.
Thus, if a state violates inequality (\ref{delta}) for at least one value of $\alpha$, it is entangled {\cite{ad}. 
Simon's version of this criterion \cite{simon} is based on expressing the partial transposition as a mirror reflection operation $\hat p_2 \to -\hat p_2$, which can be viewed as time reversal on mode 2. The PPT criterion expresses that following such a reflection, any separable state remains physical (its density operator is positive semi-definite). Conversely, an entangled  state will be detected if the corresponding reflected state is non-physical (its density operator admits a negative eigenvalue).
Remarkably, this separability condition becomes necessary {\em and} sufficient in the case of  Gaussian states and can even be extended to  ($N$+$M$)-mode Gaussian states \cite{adesso}. However, for any other state, this criterion may very often leave entanglement undetected.

Earlier work has aimed at improving the Duan-Simon separability criterion for arbitrary states. In particular, Walborn {\it et al.} \cite{walborn} reported on a separability condition using Shannon entropy, which was later extended by Huang \cite{huang}. Shchukin and Vogel \cite{shch} also derived a hierarchy of inequalities involving higher-order moments of the quadrature components (the previous criterion only depends on the first- and second-order moments). In the present work, we investigate an improvement of the Duan-Simon separability criterion that enables a stronger entanglement detection for non-Gaussian two-mode states by taking into account an additional parameter, namely, the degree of Gaussianity $g$. It is natural to expect that a stronger criterion can be obtained with more information on the state, but the additional parameter should be chosen carefully. For example, the purity of the state is not a good candidate, since the Duan-Simon criterion is necessary and sufficient for every Gaussian state of arbitrary purity. As we shall  show, the degree of Gaussianity $g$ is indeed a good choice.

In Sec.~II we start by introducing the parameter $g$ and providing some of its useful properties (more details are given in Appendix A). 
 In Sec.~III  we introduce a  criterion for a better detection of entangled states employing this degree of Gaussianity, thus improving the Duan-Simon criterion for non-Gaussian states. In Sec.~IV, we provide explicit examples of entangled non-Gaussian states that are left unnoticed by the Duan-Simon criterion, demonstrating the advantage of our  criterion (the analytical computation of parameter $g$ is detailed in Appendix B). The examples are produced from non-Gaussian states belonging both to the set of classical states (with positive Glauber P-function) and to the set of  genuinely quantum states (Fock states), reflecting the general applicability of our method.  Finally, we  conclude and discuss possible extensions of our work in Sec.~V.

\section{Degree of Gaussianity}

Gaussian states play a prominent role in continuous-variable quantum information \cite{weed}. However, several protocols necessarily require the use of non-Gaussian states, such as entanglement distillation \cite{eisert,fiurasek,giedke} or quantum error correction \cite{niset,lassen}. With the increasing importance of  non-Gaussian states, the question of measuring the Gaussian character of a state has naturally arisen. Several Gaussianity measures have been introduced (see, e.g., \cite{gauss3,gauss4,gauss5,gauss6}), but  we find it more convenient here to use the degree of Gaussianity $g$ introduced in \cite{mand}. 

Consider a two-mode state $\rho$. Its first- and second-order moments, denoted, respectively, by ${\bf d}$ and $\gamma$, are expressed from the vector of quadrature components $\hat {\bf r}=(\hat x_1,\hat p_1,\hat x_2,\hat p_2)$. The elements of the coherent vector are given by $d_j=\langle \hat r_j \rangle$, while the elements of the covariance matrix are defined as $\gamma_{ij}= \langle \hat r_i \hat r_j + \hat r_j \hat r_i \rangle - 2 d_i d_j$. We assume, with no loss of generality, that all states considered in the following have vanishing coherent vectors ($d_j=0$) since first-order moments are irrelevant as far as entanglement detection is concerned. The degree of Gaussianity $g$ of state $\rho$ is defined as
\begin{equation}
g=\frac{\mathrm{Tr}(\rho \rho^G)}{\textrm{Tr}(\rho^G\rho^G)},
\label{gauss}
\end{equation}
where  $\rho^G$ is the Gaussian state characterized by the covariance matrix $\gamma$ of state $\rho$. 

One may easily verify that $g=1$ for Gaussian states (note that the converse is not true, as shown in Appendix~\ref{annexeA}).  In addition,  the degree of Gaussianity $g$ is  invariant under Gaussian unitary transformations, transposition, and partial transposition. The proofs are provided in Appendix~\ref{annexeA}. These properties are essential for the derivations in this work. We will also exploit the fact that the knowledge of $g$ gives a tighter bound in the uncertainty relations \cite{mand}, which in turn translates into a stronger condition for detecting entanglement.

\section{Improved separability condition based on gaussianity-bounded uncertainty relations}

Let us investigate the separability of an arbitrary two-mode state $\rho$. As mentioned earlier, the PPT criterion consists in verifying the physicality of the partially transposed state $\rho^{T_2}$ (which must hold for any separable two-mode state). Then, an entangled state $\rho$ will be detected if $\rho^{T_2}$ is not physical (it exhibits a negative eigenvalue). Applying a partial transposition (acting on the second mode) in state space is equivalent to a mirror reflection $\pp\rightarrow-\pp$ in phase space.  Thus, in order to detect entanglement, we need to check the physicality of $\rho^{T_2}$ in phase space, which can be achieved based on the symplectic eigenvalues of its covariance matrix.

Suppose that the two-mode state $\rho$ has a covariance matrix $\gamma$. According to the Williamson theorem, there always exists a   unitary transformation $U_S$  mapping the state $\rho$ onto $\sigma$ such that the associated  symplectic transformation $S$  maps $\gamma$ onto 
\begin{equation}
\gamma_{\sigma}=S\gamma S^{T}=\begin{pmatrix}\nu_{+}\,\mathds{1}&0\\0&\nu_{-}\,\mathds{1}\end{pmatrix},
\label{symplectic}
\end{equation}
where $\nu_+ (\nu_-)$ is the largest (smallest) symplectic eigenvalue of $\gamma$ and $\mathds{1}$ is the $2\times 2$ identity matrix. Note that if $\rho$ is a Gaussian state, then $\sigma$ is a tensor product of two thermal states; otherwise $\sigma$ is a two-mode non-Gaussian state that has the same covariance matrix $\gamma_{\sigma}$.
The uncertainty principle implies that the inequality  $\nu_+ \geq\nu_-\geq1$ must be respected for any physical state \cite{weed}. Applying this condition to the partially transposed state $\rho^{T_2}$, we understand that the entanglement of states $\rho$ is detected whenever the smallest symplectic eigenvalue of $\rho^{T_2}$ is strictly smaller than 1, which is the core of the Duan-Simon criterion.

Let us now introduce our improved criterion. In order to detect the entanglement of state $\rho$, we apply a partial transposition  on the second mode (which may lead to an unphysical state) followed by a symplectic transformation, which gives access to the symplectic eigenvalues $\nu_\pm $ of the partially transposed state $\rho^{T_2}$. This is the entanglement analyzing box shown in Fig. \ref{box}.

    \begin{figure}[h!]
\includegraphics[trim=8cm 21cm 1cm 2.5cm , clip,width=0.9\columnwidth]{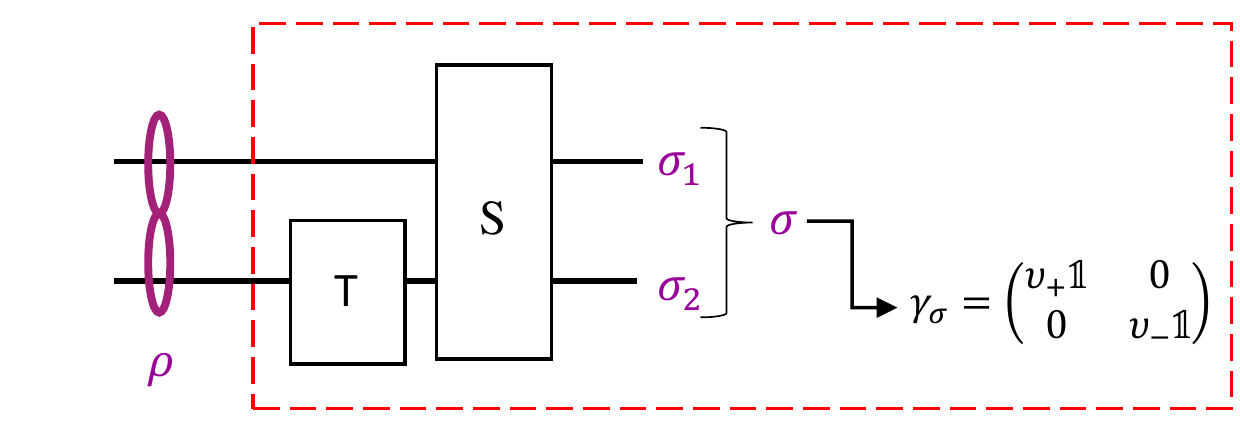}
\caption{\label{box} Entanglement analyzing box. A partial transposition T  and symplectic transformation S are applied to state $\rho$, giving access to the symplectic eigenvalues $\nu_\pm $ of the partial transposed state. Note that this circuit is not physical since T is antiunitary.}
\end{figure}

A key observation is that the inequality $\nu_+~ \geq~\nu_-~\geq~1$  boils down to expressing the uncertainty relation for the two modes making $\sigma$ at the output of the entanglement analyzing box. Indeed, we have $\textrm{det}(\gamma_{\sigma_{1}})= (\nu_{+})^2 \ge 1$ and  $\textrm{det}(\gamma_{\sigma_{2}})= (\nu_{-})^2 \ge 1$, where $\gamma_{\sigma_1}$ ($\gamma_{\sigma_2}$) refers to the covariance matrix of the first (second) mode of $\sigma$.  
Furthermore, a tighter lower bound on the uncertainty $\textrm{det}(\gamma)$ of a mode can be obtained if the degree of Gaussianity $g$ of this mode is known \cite{mand} [we use definition (\ref{gauss}) for a single mode]. 
Combining these elements,  we can detect the non-physicality of $\rho^{T_2}$ whenever the lowest symplectic eigenvalue $\nu_-$ lies under the lower bound corresponding to the degree of Gaussianity $g_2$ of $\sigma_2$, as shown in Fig.~\ref{graphkat}. 
This lower bound is equal to $1$ for $g_2=1$, but is strictly larger than $1$ for non-Gaussian states with $g_2\ne 1$.

\begin{figure}[h!]
\includegraphics[width=0.6\columnwidth]{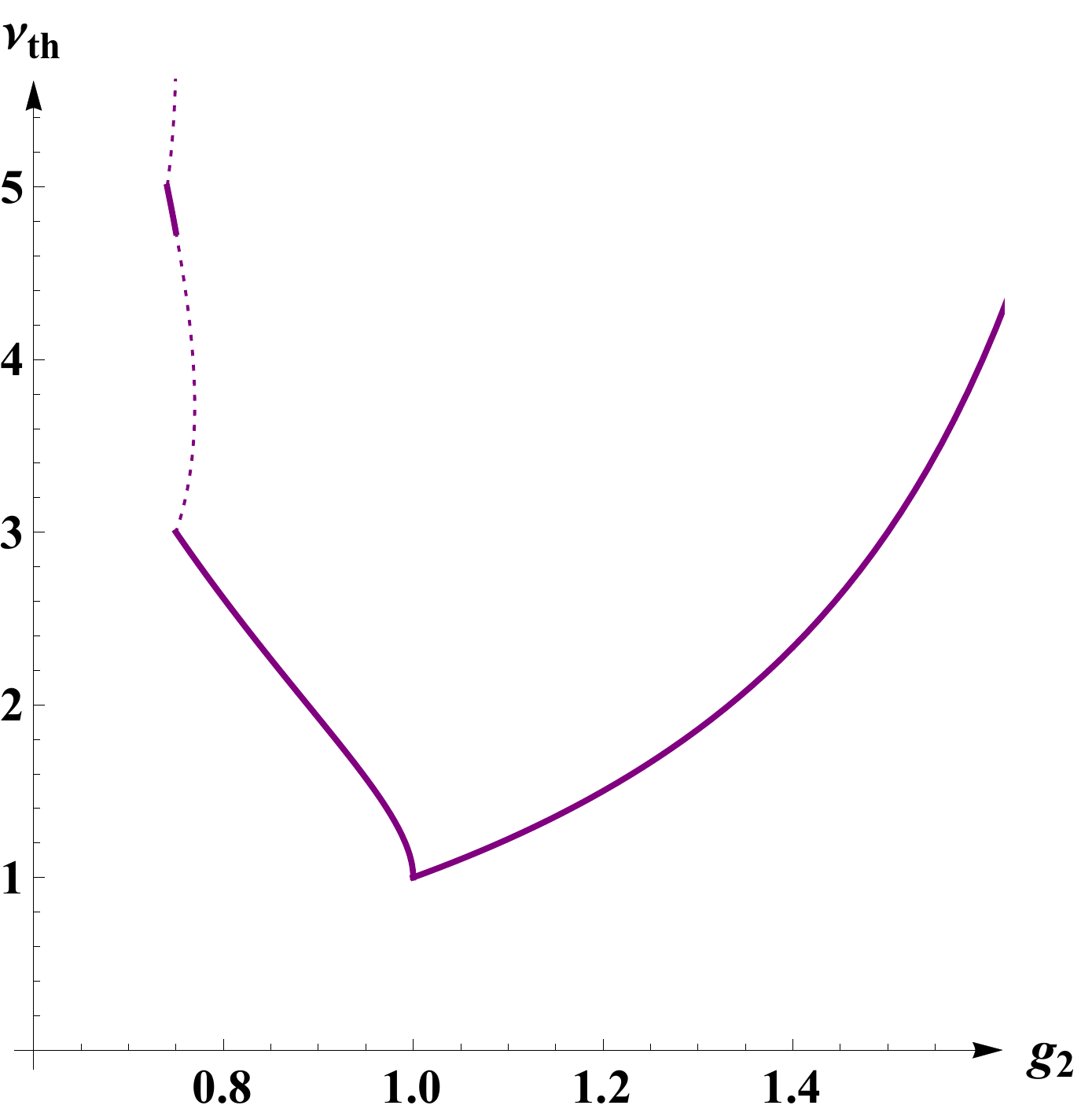}
\caption{\label{graphkat} Plot of $\nu_{\mathrm{\mathrm{th}}}$, the threshold (minimum allowed) value for $\nu_{-}$, as a function of the degree of Gaussianity $g_2$ (details are given in \cite{mand}). All physical states lie on or above this curve. Note that for $g_2<1$, the curve exhibits some discontinuities (see main text).} 
\end{figure}

Hence, we obtain an improved separability criterion that works as follows. After applying the entanglement analyzing box of Fig.~\ref{box} to state $\rho$, we detect its entanglement if the symplectic eigenvalue of the reduced state $\sigma_2$ is smaller than a bound, which is a function of the degree of Gaussianity $g_2$ of $\sigma_2$. In other words, 
\begin{equation}
\nu_-<\nu_{\mathrm{\mathrm{th}}}(g_2)\quad\Rightarrow\quad\text{entanglement},
\label{improved-criterion}
\end{equation}
where $\nu_{\mathrm{\mathrm{th}}}(g_2)$ is the threshold given by the curve in Fig.~\ref{graphkat}.
 If $g_2\geq1$, the curve is given by $\nu_{\mathrm{\mathrm{th}}}= g_2/(2 - g_2)$. If $g_2<1$, the parametric equations of the curve are given by 
 \begin{eqnarray}
 \nu_{\mathrm{\mathrm{th}}}&=&2 n + 3 - 2 r,\nonumber\\
  g_2&=&\frac{2 \nu_{\mathrm{\mathrm{th}}}  (\nu_{\mathrm{\mathrm{th}}} -1)^n }{(\nu_{\mathrm{\mathrm{th}}} +1)^{n+1}}\left(\frac{(\nu_{\mathrm{\mathrm{th}}} -1) (1-r)}{\nu_{\mathrm{\mathrm{th}}} +1}+r\right),
 \end{eqnarray}
  where $n\in\mathbb{N}$ and $r\in[0,1[$. The latter curve consists of consecutive segments, each corresponding  to a binary mixture of nearest-neighbor Fock states $\ket{n}$ and $\ket{n+1}$. In the examples that we will discuss in Sec. IV, the degree of Gaussianity always lies in the segment where $n=0$, which corresponds to
  \begin{equation}
  \nu_{\mathrm{\mathrm{th}}}(g_2)=  \frac{2-g_2+2 \sqrt{1-g_2}}{g_2}\qquad\text{for }\,\,\,3/4 \le g_2 \le 1.
  \label{nuth}
\end{equation}

 In order to exploit condition (\ref{improved-criterion}), the last step is thus to  compute the degree of Gaussianity $g_2$ as given by Eq.~(\ref{gauss}). The analytical computation of $g_2$ is not trivial for an arbitrary two-mode state (although we give an explicit method for some class of states in Appendix B), but at least a numerical computation is always feasible based on the Wigner function. First, we remark that the denominator of $g_2$ is simply equal to $1/\nu_-$ since it corresponds to the purity of a Gaussian state [see Eq.~(\ref{denom}) in Appendix \ref{AnnexeB}]. To express the numerator of $g_2$, we use the Wigner function $\tilde{W}_2(x_2,p_2)$ of the second mode $\sigma_2$ at the output of the entanglement analyzing box. Starting from $W(x_1,p_1,x_2,p_2)$, namely, the Wigner function of the initial two-mode state $\rho$, we find $W^{T_2}(x_1,p_1,x_2,p_2)=W(x_1,p_1,x_2,-p_2)$ after partial transposition and then  $\tilde{W}(x_1,p_1,x_2,p_2)$ after symplectic transformation, corresponding to a change of variable $\vec{r} \rightarrow S\vec{r}$. Finally, we integrate over $x_1$ and $p_1$ to have the Wigner function of the second mode $\sigma_2$, which gives
\begin{equation}
g_2=\frac{\tr(\sigma_2\sigma_2^G)}{\tr(\sigma_2^G\sigma_2^G)}=2\pi\nu_-\,\int\tilde{W}_2(x_2,p_2)\tilde{W}_2^G(x_2,p_2)dx_2dp_2
\end{equation}
where $\tilde{W}_2^G(x_2,p_2)$ is the Wigner function of the Gaussian state with covariance matrix $\nu_-\mathds{1}$.

 \begin{figure}[h!]
\begin{center}
\includegraphics[trim=3cm 0cm 0cm 0cm , clip,width=0.7\columnwidth]{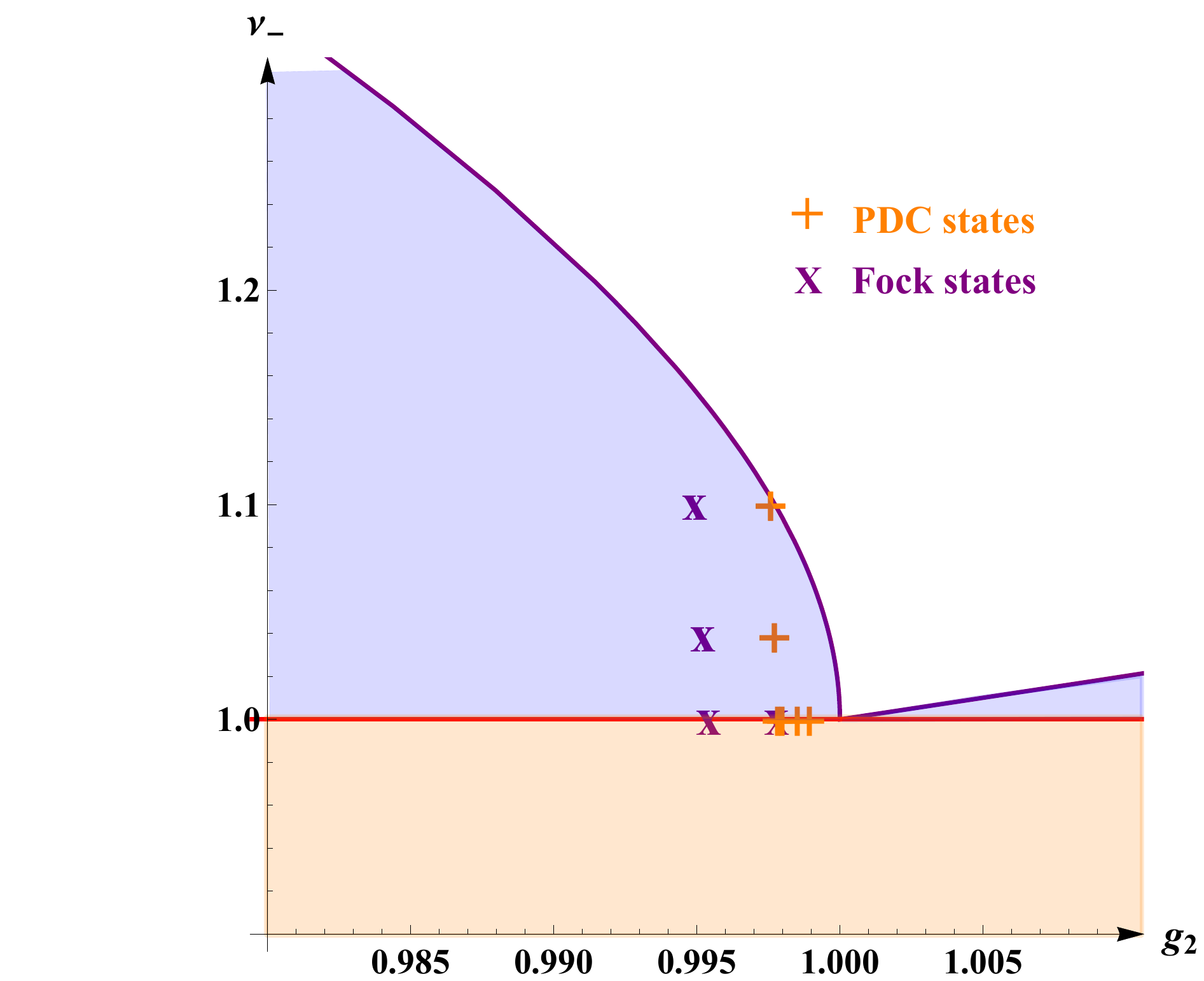}
\caption{\label{result} 
Examples of non-Gaussian entangled states generated from Fock ($\times$)  or  phase-diffused coherent (PDC)  (+) states, which are detected by our improved criterion, but not otherwise. In general, all entangled states detected by the Duan-Simon criterion $(\nu_- \ge 1)$ lie in the red (lower) zone, while the entangled states that are detected by our criterion but remain undetected by the Duan-Simon criterion lie in the blue (intermediate) zone. The white (upper) zone contains either separable or undetected entangled states. The curve $\nu_{\mathrm{\mathrm{th}}}(g_2)$ separating the blue and white zones corresponds to the lower bound on $\nu_-$ for a fixed degree of Gaussianity $g_2$ (see Fig. \ref{graphkat}).
  }
\end{center}
\end{figure}

Figure \ref{result} enables us to visualize how entanglement detection is improved by our method. Three  distinct zones are represented, delimited by the curve of Fig. \ref{graphkat} and by the constant line $\nu_-=1$.
 If a state lies in the red (lower) zone, it is an entangled state that is detected by the Duan-Simon criterion ($\nu_-<1$), hence it is uninteresting for our purposes here. If it lies in the white (upper) zone, no conclusion can be made because the partially transposed state is physical. However, interestingly, if it lies in the blue (intermediate) zone, we detect entanglement which was otherwise unnoticed.

We remark that, since partial transposition and symplectic transformation conserve the Gaussian character of a state (see Appendix \ref{annexeA}), if $\rho$ is a Gaussian state, then $\sigma$ and the reduced states $\sigma_1$ and $\sigma_2$ are also Gaussian. Then $\nu_{\mathrm{\mathrm{th}}}(1)=1$, and we recover the (necessary and sufficient) Duan-Simon separability criterion for Gaussian states, as expected.}

 Let us also mention that our criterion does not improve entanglement detection when the covariance matrix of $\rho$ is diagonal, since the partially transposed state then necessarily remains physical. For example, the ``NOON'' states of the form $(\ket{N0}+\ket{0N})/\sqrt{2}$ have a diagonal covariance matrix for $N\geq2$. Thus, even though those states are always entangled, we cannot do any better than the Duan-Simon criterion, and entanglement is undetected by our criterion. However, there exist many other interesting cases where our method is useful, as shown in the next section. 
   
\section{Examples of non-Gaussian entangled states detected by the improved criterion}

In this section, we apply our  criterion to two types of non-Gaussian states. Those examples have in common that entanglement is not detected on the sole basis of the covariance matrix (using the Duan-Simon criterion), but is detected  exploiting the degree of Gaussianity.

\subsection{Non-Gaussian states generated from Fock states or phase-diffused coherent states}

\begin{figure}[h!]
\begin{center}
\includegraphics[trim=5cm 12.5cm 5cm 10.5cm , clip,width=0.8\columnwidth]{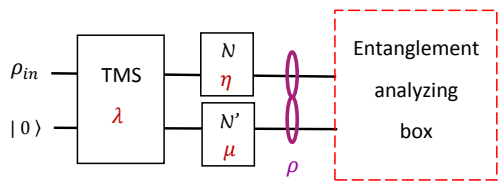}
\caption{\label{Figure}Quantum circuit used to prepare the non-Gaussian states $\rho$ with covariance matrix $\gamma$. The entanglement of $\rho$ is analyzed by the red box (see Fig. \ref{box}), which pictures our improved criterion. } 
\end{center}
\end{figure}

The first example uses non-Gaussian states as generated by the circuit of Fig. \ref{Figure}. 
The preparation of the states works as follows. Initially, we have a Fock-diagonal state $\rho_{\textrm{in}}=\sum_{n=0}^{\infty} \phi_n\ket{n}\bra{n}$ of covariance matrix $\gamma_{\mathrm{in}}=a\, \mathds{1}$ in the first mode, where $a \in [1,\infty)$, and the vacuum state in the second mode. Both states are processed through a two-mode squeezer (TMS) of parameter $\lambda \in [0,1)$. Note that if $\rho_{\mathrm{in}}$ is a Fock state $\ket{n}$, at this point of the circuit we have a photon-added EPR state, which we know is always entangled (its entanglement is monotonically increasing with $n$)  \cite{nava}. Of course, if we have the vacuum on both modes, the resulting state is simply an EPR state. The next step in the circuit consists in 
processing each mode of the state through two independent Gaussian additive-noise channels $\mathcal{N}$ and $\mathcal{N}'$.
The variance of the added noise on the first (second) mode is denoted by $\eta$ ($\mu$). This  construction ensures that the resulting (not necessarily Gaussian) state will always be physical provided  $a\ge 1$, $0\le \lambda <1$, and $\eta,\mu \ge 0$.

The  state $\rho$ at the output of this circuit has a covariance matrix
\begin{equation}
\label{gammaref}
\gamma=\begin{pmatrix}
   \Big(   \frac{a+\lambda^2}{1-\lambda^2}+\eta  \Big)\,\,\, \mathds{1} &\frac{(a+1)\lambda}{1-\lambda^2} \,\sigma_z  \\\frac{(a+1)\lambda}{1-\lambda^2}\, \sigma_z&    \Big( \frac{a\lambda^2+1}{1-\lambda^2}+\mu    \Big) \, \mathds{1}
\end{pmatrix},
\end{equation}
where $\sigma_z$ is the third Pauli matrix. 
Note that this form for a covariance matrix is actually quite general. Indeed, 
Duan {\it et al.} have shown \cite{duan} that any covariance matrix of a two-mode state can be transformed into the form
\begin{equation}
\gamma=\begin{pmatrix}n&&c&\\&n&&d\\c&&m&\\&d&&m\end{pmatrix}
\label{covmat}
\end{equation}
by applying local linear unitary Bogoliubov operations, i.e., combinations of squeezing transformations and rotations. These operations do not influence the separability of the state, and are thus always allowed when studying entanglement. The covariance matrix (\ref{gammaref}) depends on three parameters while the most general form (\ref{covmat}) has only one additional parameter, which reflects that (\ref{gammaref}) encompasses a wide class of two-mode Gaussian states.

The entanglement of the resulting state $\rho$ is now analyzed 
as depicted by the red box of Fig. \ref{box}. The resulting state $\sigma$ has a covariance matrix of the form of Eq.~(\ref{symplectic}) and the symplectic eigenvalues of $\rho^{T_2}$ can be expressed as a function of the different parameters characterizing $\gamma$, namely
\begin{eqnarray}
\nu_{\pm}&=&\frac{1}{2}\Bigg(       \frac{(a+1)(1+\lambda^2)}{1-\lambda^2}+ \eta+\mu\nonumber\\
& & \pm\sqrt{   (a-1+\eta-\mu)^2+\frac{4(a+1)^2\lambda^2}{(1-\lambda^2)^2}    }     \,    \Bigg).
\end{eqnarray}
Note that this expression is valid regardless of whether $\rho$ is Gaussian or not.
At this point, using the Duan-Simon separability criterion (ignoring whether $\rho$ is Gaussian or not)  would detect an entangled state only if $\nu_-<1$. 
However, we can improve on this by taking into account the degree of Gaussianity $g_2$ at the output of the circuit of Fig. \ref{box}, as explained previously. The calculation of $g_2$ could be done through the computation of the Wigner function, but this would require some numerical integrations. In Appendix \ref{AnnexeB}, we give a different way of calculating $g_2$ for this specific example. The final expression is not very elegant, but calculations are performed completely analytically.

Now that we have defined a circuit to generate families of non-Gaussian states and detect their entanglement, we will focus on some explicit examples of such states in order to illustrate the usefulness of our improved criterion. In the circuit of Fig. \ref{Figure}, we start with Fock-diagonal states $\rho_{\textrm{in}}$, which have a diagonal covariance matrix $\gamma_{\textrm{in}}$ with variance  $a=\sum_{n=0}^{\infty}\phi_n(2n+1)$.  We are interested in non-Gaussian states $\rho_{\textrm{in}}$ and will consider two rather extreme cases of such states. The first case is a single Fock state $\ket{n}$ with $n>0$, the parameter of the covariance matrix being thus $a=2n+1$. This state has clear quantum features, such as negative parts in the Wigner function. Our second choice is a non-Gaussian mixture of coherent states with a random phase, which can be viewed as ``classical''. This phase-diffused coherent (PDC) state can equivalently be represented as a mixture of Fock states following a Poisson distribution
   \begin{equation}
   \rho_{\textrm{in}}=\sum_{k=0}^{\infty} e^{-(a-1)/2}\frac{\left(\frac{a-1}{2}\right)^k}{k!}\ket{k}\bra{k}.
   \end{equation}
   These two examples for $\rho_{\textrm{in}}$ are simple at a theoretical level and may also be implemented experimentally. 
   For the state $\rho$ to be feasible experimentally, we will focus on values of the parameter $\lambda$ of the two-mode squeezer that are smaller than 0.8 ($\approx10$ dB). The values of the noise variances $\eta$ and $\mu$ will be chosen smaller than 2 units of vacuum noise because otherwise the state $\rho$ is necessarily separable (regardless of whether it is Gaussian or not). Indeed, each mode of  $\rho$  can be seen as the output of a classical Gaussian additive noise channel and it is known that such a channel is entanglement breaking if $\eta\geq2$ ($\mu\geq2$) \cite{holevo}.

In Fig. \ref{result} we exhibit explicit examples of non-Gaussian states $\rho$ that are generated from $\rho_{\textrm{in}}$ being either a Fock state or a phase-diffused coherent state. The corresponding numerical values of the circuit parameters ($a,\lambda,\mu,\eta$) are displayed in Table \ref{result_tableau}. We first choose sets of values of the circuit parameters such that $\nu_-=1$, implying that the Duan-Simon criterion does not detect entanglement. In this case, entanglement is detected as soon as $g_2\neq1$, so all these example states are proven to be entangled with our improved separability criterion. We then extend our search to larger values of $\nu_-$. An entangled state is then detected whenever $\nu_-<\nu_{\mathrm{th}}(g_2)$. Since in our examples $3/4\le g_2 < 1$, the function $\nu_{\mathrm{th}}(g_2)$ is given by Eq. (\ref{nuth}).
  All points localized in the blue zone are thus examples of non-Gaussian entangled states that are detected by our improved separability criterion but not otherwise.  We remark that entangled states can be found with both choices of $\rho_{\textrm{in}}$ (either a highly non-classical Fock state or a classical mixture of phase-diffused coherent states).

\begin{table}[h!]
\begin{center}
\begin{tabular}{c|c|c|c|c|c|c} 
   \hline\hline
type of $\rho_{\mathrm{in}}$&  $a$&$\lambda$&$\eta$&$\mu$&$\nu_-$&$g_2$ \\
  \hline\hline
  
Fock&  3&0.6&1/13&1&1&0.99541\\
Fock&3&0.3&0.1&228/757&1&0.99780\\
Fock& \hspace{1pt} 3 \hspace{1pt}& \hspace{1pt}0.5 \hspace{1pt}& \hspace{1pt}0.1 \hspace{1pt}& \hspace{1pt}0.9 \hspace{1pt}& \hspace{1pt}1.1 \hspace{1pt}& \hspace{1pt}0.99492 \hspace{1pt}\\
   Fock& 3&0.5&0.1&0.8&1.04&0.99521\\

PDC&2&0.3&0.1&513/1271&1&0.99798\\
PDC&2&0.7&0.1&931/677&1&0.99850\\
PDC&3&0.6&1/13&1&1&0.99781\\
PDC&3&0.3&0.1&228/757&1&0.99893\\

PDC&  3&0.5&0.1&0.9&1.1&0.99758\\
PDC&    3&0.5&0.1&0.8&1.04&0.99771\\

    \hline
\end{tabular}
\caption{\label{result_tableau} Values of the circuit parameters used to generate the examples of non-Gaussian entangled states that are detected by our improved separability  criterion. The corresponding values of $\nu_-$ and $g_2$ are also given.}
\end{center}
\end{table}
   
   \begin{figure}[h!]
\begin{center}
\includegraphics[trim=18cm 0cm 0cm 0cm , clip,width=0.6\columnwidth]{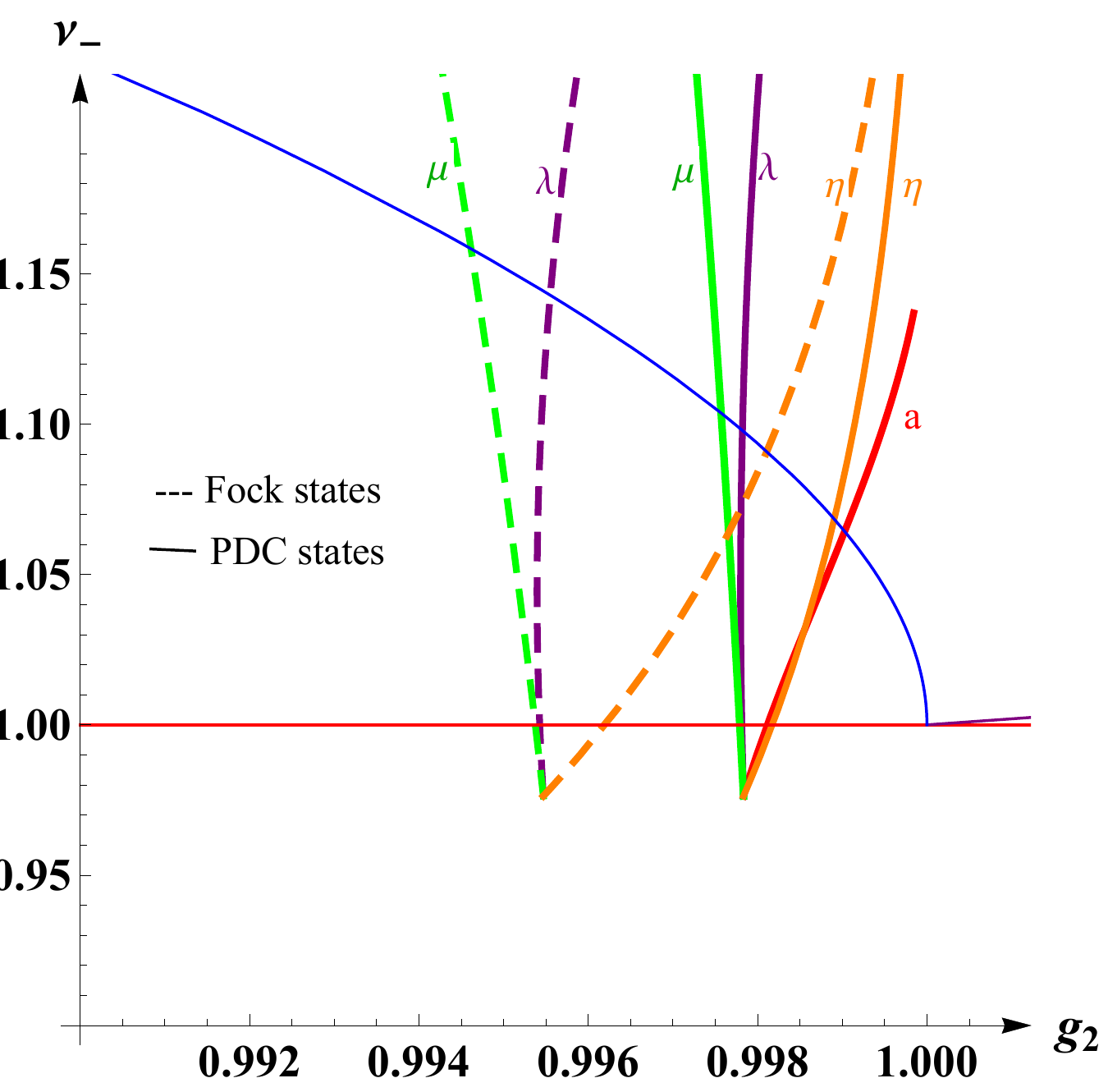}
\caption{\label{result_2} Evolution of symplectic eigenvalue $\nu_-$ and degree of Gaussianity $g_2$ when varying the circuit parameters, illustrating how they influence the separability of the states. For each curve, one of the parameters varies while the others are fixed, starting from $a=3,\,\lambda=0.5,\,\eta=0.1$ and $\mu=0.7$ (at the bottom of the curve). Increasing $\nu_-$ is achieved by increasing $a$, $\eta$, and $\mu$ and decreasing $\lambda$.  The blue curve  is given by $\nu_{\mathrm{\mathrm{th}}}(g_2)$.}
\end{center}
\end{figure}
   
Interestingly, for all states created with our circuit when Gaussian noise is added on the first mode only (i.e., $\mu=0$), we find out that the Duan-Simon separability criterion becomes necessary and sufficient, even for non-Gaussian states. Indeed, if $\eta<2$, the symplectic eigenvalue $\nu_-$ is smaller than 1 for all values of $a$ and $\lambda$, hence the state $\rho$ is entangled. In contrast, if $\eta\geq2$, we have an entanglement breaking channel, so we know that the state $\rho$ is necessarily separable. This confirms the validity of our method.

Finally, Fig. \ref{result_2} illustrates how the different circuit parameters influence the separability of the state. Starting with a Fock state (or with a phase-diffused coherent state) with circuit parameters $a=3,\,\lambda=0.5,\,\eta=0.1$, and $\mu=0.7$ (the lowest points of the curves), we see that by varying one of the parameters we can always create entangled states that are unnoticed by the Duan-Simon criterion. (Note that there is no curve to plot corresponding to varying $a$ for Fock states since $a$ can only take odd integer values in this case.)


\subsection{Squeezed single-photon path-entangled state}

\begin{figure}[h!]
\begin{center}
\includegraphics[trim=5cm 4cm 5cm 19cm , clip,width=0.8\columnwidth]{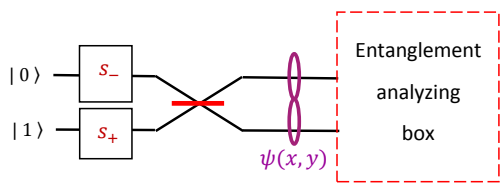}
\caption{\label{path}Quantum circuit used to prepare a squeezed single-photon path-entangled state $\psi(x,y)$. The entanglement of the state is analyzed by the red box (see Fig. \ref{box}). } 
\end{center}
\end{figure}

As a second example, let us consider a squeezed single-photon path-entangled state, i.e. the  non-Gaussian state created from the circuit of Fig. \ref{path}. 

A vacuum and single-photon Fock states are both squeezed, with respective squeezing parameters $s_-$ and $s_+$, and are then coupled with a balanced beam splitter. The wave function of the output (pure) state has the form \cite{walborn}
\begin{equation}
\psi(x,y)=\frac{(x+y)}{\sqrt{\pi s_-s_+^3}}e^{-\frac{(x+y)^2}{4s_+^2}-\frac{(x-y)^2}{4s_-^2}}.
\end{equation}
This state is obviously entangled for all values of $s_{\pm}$, but the Duan-Simon separability criterion detects entanglement only for $s_-/s_+>\sqrt{3}$ or $s_-/s_+<1/\sqrt{3}$. However, similarly as what  Walborn {\it et al.}  \cite{walborn} have shown using their entropic entanglement criterion, we can detect entanglement for all values of $s_{\pm}$ with our improved criterion.
Let us suppose that  $s_-/s_+\geq1$. Applying the ``entanglement analyzing box'' to this state, we find that
\begin{equation}
\nu_-=\sqrt{3}\,\frac{s_+}{s_-} \quad\text{and}\quad g_2=\frac{3}{4}\sqrt{\frac{3}{2}},
\end{equation}
where $g_2$ is computed with the help of Wigner functions.
Therefore, according to Eq. (\ref{nuth}),  $\nu_{\mathrm{\mathrm{th}}}(g_2)=1.7986$ and entanglement is detected if 
\begin{equation}
\nu_-<\nu_{\mathrm{\mathrm{th}}}(g_2)\quad\Leftrightarrow\quad\frac{s_-}{s_+}>0.963.
\end{equation}
However, we supposed at the beginning that $s_-/s_+\geq1$. Entanglement is thus always detected.
The same analysis can be done if $s_-/s_+<1$.
Although the entropic criterion of Walborn {\it et al.} also detects the entanglement of this state for all $s_{\pm}$, we believe that our method is easier to apply.


\section{Conclusion}

In this paper we have introduced a  continuous-variable separability criterion exploiting the degree of Gaussianity of the state, thereby allowing a stronger detection of two-mode non-Gaussian entangled states. Our criterion works by verifying the physicality of the symplectic eigenvalues of the partially transposed state  in terms of Gaussianity-bounded uncertainty relations. 
We demonstrated the advantages of our method by providing  explicit examples of states whose entanglement is detected by our criterion but left undetected by the Duan-Simon criterion based on the EPR variance only. We proposed an optical circuit for creating a family of such states and studied the entanglement detection as a function of the parameters of the circuit. The values of those circuit parameters were chosen so that these example states could be experimentally generated to demonstrate the method.
The general applicability of the method is witnessed by the fact that these example states can be generated both from genuinely quantum non-Gaussian states (Fock states) and from classical non-Gaussian mixtures of phase-diffused coherent states (states with a positive P-function). We expect that many more examples of entangled states could be found, first by testing different values of the parameters $a,\lambda,\eta$ and $\mu$, second by generalizing the circuit (for example, at the very beginning of the circuit, one can insert a thermal state instead of the vacuum), or simply by devising a new circuit  generating other types of non-Gaussian states such as those of our second example.

 As mentioned in the Introduction, a separability condition such as inequality (\ref{delta}) cannot be rewritten  with a tighter lower bound that would solely depend on purity ${\rm Tr}(\rho^2)$. This is because the Duan-Simon criterion is necessary and sufficient for all Gaussian states (of arbitrary purity).  Hence, the lower bound in inequality (\ref{delta}) cannot be moved upward without being violated by some mixed Gaussian states that are known to be separable.  However, we expect that our separability criterion may be further improved by taking into account both the degree of Gaussianity and purity of the state, and then making use of the purity- and Gaussianity-bounded uncertainty relations \cite{mand2}. This topic is worth further investigation.

\section{Acknowledgments}
This work was supported by the F.R.S.-FNRS Foundation under Project No. T.0199.13 and by the Belgian Federal IAP program under Project No. P7/35 Photonics@be. A. H. acknowledges financial support from the F.R.S.-FNRS Foundation, and A. M. acknowledges financial support from the Ministry of Education and Science of the Republic of Kazakhstan (Contract No. 339/76-2015).
\vspace{-5pt}

\appendix

\section{Properties of the degree of Gaussianity}
\label{annexeA}

The degree of Gaussianity $g$  defined by Eq.~\eqref{gauss}
holds for an $N$-mode state (here, we focus on $N=1$ and $N=2$). It obeys the
following properties : 
\begin{itemize}
\item For a Gaussian state  $\rho=\rho^G$ this measure obviously gives $g=1$.
However, $g=1$ does not imply that state $\rho$ is necessarily  Gaussian. Let us present some counter examples. 
Consider $\rho$ being the mixture of two Fock states:
\begin{equation}
\rho=\frac{1}{2\sqrt{2}}\ket{2}\bra{2}+\left(1-\frac{1}{2\sqrt{2}}\right)\ket{0}\bra{0}.
\end{equation}
The covariance matrix of $\rho$,
\begin{equation}
\gamma=\begin{pmatrix}a&&0\\0&&a\end{pmatrix}\qquad\text{with}\qquad a=1+\sqrt{2},
\end{equation}
determines a Gaussian (thermal) state
\begin{equation}
\rho^G=\frac{\sqrt{2}}{1+\sqrt{2}}\sum_j\left(\frac{1}{1+\sqrt{2}}\right)^j\ket{j}\bra{j}.
\end{equation}
It is then easy to see that
\begin{equation}
\mathrm{Tr}[\rho^G\rho^G]=\mathrm{Tr}[\rho\rho^G]=\frac{1}{1+\sqrt{2}}.
\end{equation}
This obviously gives $g=1$, although $\rho$ is a non-Gaussian state.
Other counter examples of non-Gaussian states with $g=1$
may be found among the states of the form
\begin{equation}
\rho=p\,\ket{n}\bra{n}+\left(1-p\right)\ket{0}\bra{0}.
\end{equation}
Given $n$, the real roots of the equation
\begin{equation}
\label{rootsp}
\qquad\,\,\,(1+2np)(np)^n-(1+2np-n)(1+np)^n=0
\end{equation}
satisfying  $0<p< 1$ provide $g=1$.
Note that Eq.~\eqref{rootsp} is a polynomial of degree $n$, therefore, the number of its roots providing counter examples is expected to increase with $n$. 

\item The degree of Gaussianity $g$ is invariant under Gaussian unitary operations.

{\it Proof:} 
Consider an arbitrary state $\rho$ and corresponding Gaussian state $\rho^G$.  A Gaussian unitary operator $U^G$ transforming Gaussian states to Gaussian states transforms $\rho$ to  $\rho' = U^G \rho (U^G)^\dag$. The same operator similarly transforms the Gaussian state $\rho^G$  to  $\rho'^G = U^G \rho^G (U^G)^\dag$. The later transformation is equivalent to a symplectic transformation of the corresponding covariance matrices. By construction of $\rho^G$, its covariance matrix is also the covariance matrix of $\rho$. This covariance matrix  is transformed by the symplectic transformation into the covariance matrix of $\rho'^G$. Therefore,  $\rho'^G$ is the Gaussian state corresponding to $\rho'$. Then a simple calculation gives us the desired result:
\begin{eqnarray}
g'&=&\frac{\textrm{Tr}(\rho' \rho'^G)}{\textrm{Tr}(\rho'^G\rho'^G)}\nonumber \\
  & = & \frac{\textrm{Tr}(U^G\rho (U^G)^{\dag} U^G\rho^G(U^G)^{\dag} )}{\textrm{Tr}(U^G\rho^G(U^G)^{\dag} U^G\rho^G(U^G)^{\dag}) }\nonumber\\
&=&\frac{\textrm{Tr}(\rho \rho^G)}{\textrm{Tr}(\rho^G\rho^G)}=g,
\end{eqnarray}
where we used the invariance of the trace under cyclic permutations. $\blacksquare$
\item The degree of Gaussianity $g$ is invariant under  partial transposition.

{\it Proof:}  
Partial transposition implies sign-flip
of one of the two momentum quadratures (say  $p_2\rightarrow-p_2$), i.e. one of the arguments of the Wigner function 
describing a two-mode state. Then we have
\begin{widetext}
\begin{eqnarray}
g'=\frac{\textrm{Tr}(\rho' \rho'^G)}{\textrm{Tr}(\rho'^G\rho'^G)} 
  & =  & \frac{(2\pi)^2\int dx_1 dp_1 dx_2 dp_2 W_{\rho'}(x_1,p_1,x_2,p_2)W_{\rho'^G}(x_1,p_1,x_2,p_2)}{(2\pi)^2\int dx_1 dp_1 dx_2 dp_2 W_{\rho'^G}(x_1,p_1,x_2,p_2)W_{\rho'^G}(x_1,p_1,x_2,p_2)}\nonumber\\
&=&\frac{(2\pi)^2\int dx_1 dp_1 dx_2 dp_2 W_{\rho}(x_1,p_1,x_2,-p_2)W_{\rho^G}(x_1,p_1,x_2,-p_2)}{(2\pi)^2\int dx_1 dp_1 dx_2 dp_2 W_{\rho^G}(x_1,p_1,x_2,-p_2)W_{\rho^G}(x_1,p_1,x_2,-p_2)}=\frac{\textrm{Tr}(\rho\rho^G)}{\textrm{Tr}(\rho^G\rho^G)} =g,
\end{eqnarray}
\end{widetext}
where  at the last step we changed the variables as   $-p_2\rightarrow p_2$.  $\blacksquare$

\item The degree of Gaussianity $g$ is invariant under transposition.

{\it Proof:} The proof follows the same steps as for the partial transposition, but in this case, we have both $p_2\rightarrow-p_2$, and $p_1\rightarrow-p_1$. This does not change the conclusion. $\blacksquare$
 
  \end{itemize}

Finally, we would like to comment on the possibility of an experimental estimation of the degree of Gaussianity $g$. Following   \cite{mand}, an expression of $g$ 
as a converging series on the radial  moments $\left\langle r^{2n+1}\right\rangle$ of the Wigner function is derived
for states with an angular-independent Wigner function and covariance matrix of the form $\gamma=a\mathds{1}$:
\begin{equation}
g=4\pi\sum_{n=0}^{\infty}\frac{(-1)^n\left\langle r^{2n+1}\right\rangle}{n!a^n}\label{se}.
\end{equation}
 Equation (\ref{se}) is also applicable to
states with a phase-dependent Wigner function but in this case one needs to employ 
the phase-averaged quantity $\frac{1}{2\pi}\int_0^{2\pi} \left\langle r^{2n+1}\right\rangle_\varphi d\varphi$  instead of $\left\langle r^{2n+1}\right\rangle$.  Equation (\ref{se}) gives evidence that the experimental estimation of the degree of Gaussianity is a feasible task for states where the  higher moments are of decreasing strength  (see, for example, Ref.~\cite{gaussexp}).


\section{Computation of the degree of Gaussianity}
\label{AnnexeB}

Let us show how one can perform the computation of the degree of Gaussianity $g_2$ of the reduced one-mode state corresponding to $\sigma_2$ at the output of the ``entanglement analyzing box'' of Fig.~\ref{box} (this state corresponds to the smallest symplectic eigenvalue $\nu_-$). The same technique allows computation of the degree of Gaussianity $g$ of the  two-mode state $\rho$ (see Fig.~\ref{Figure}) as well.

By our convention the Gaussian state $\sigma_2^G$ has the same covariance matrix 
$\gamma_{\sigma_2}=\nu_-\mathds{1}$ as $\sigma_2$. Then, since it corresponds to the purity of $\sigma_2^G$, the denominator in the definition of  $g_2$ given by  Eq.~\eqref{gauss} is trivially evaluated as 
\begin{equation}\label{denom}
\textrm{Tr}[\sigma_2^G\sigma_2^G]=\frac{1}{\sqrt{\det(\gamma_{\sigma_2})}}=\frac{1}{\nu_-}.
\end{equation}

 The evaluation of the numerator in Eq.~\eqref{gauss} is more involved. Although after the two-mode squeezer, we have a very simple form of the density matrix of the state $\rho_{\mathrm{TMS}}=U_{\lambda}(\rho_{\mathrm{in}}\otimes\ket{0}\bra{0})U_\lambda^{\dag} $,  the addition of the Gaussian noises makes the density matrix of state $\rho$  (and so the ones of the reduced states $\sigma_1$ and $\sigma_2$) very hard to express in a simple form. State $\rho$  is obtained as a result of the application of Gaussian additive noise channels $\Phi_\eta$ and $\Phi_\mu$ to the first and second modes of $\rho_{\mathrm{\mathrm{TMS}}}$ correspondingly 
   \begin{eqnarray}
\rho&=&(\Phi_\eta\otimes\Phi_\mu)[\rho_{\mathrm{\mathrm{TMS}}}] \label{rhonoise} \\
     & = &  \int dx_1dp_1dx_2dp_2 \,\,\frac{1}{2\pi \eta}e^{-\frac{x_1^2+p_1^2}{2\eta}}\frac{1}{2\pi \mu}e^{-\frac{x_2^2+p_2^2}{2\mu}}\nonumber\\
&\times &D(x_1,p_1)D(x_2,p_2)\,\rho_{\mathrm{\mathrm{TMS}}}\, D^{\dag}(x_1,p_1)D^{\dag}(x_2,p_2), \nonumber
\end{eqnarray}
where $D(x,p)$ is the displacement operator.
We will perform calculations avoiding the  direct use of the density matrix $\sigma_2$.  Instead we will use the following construction. Let $\sigma$ be the density matrix of the two-mode state that has $\sigma_1$ and $\sigma_2$ as reduced states. Then we have the following equivalent representation of the numerator in Eq. \eqref{gauss}:
   \begin{eqnarray}
\textrm{Tr}[\sigma_2\sigma_2^G]&=&\textrm{Tr}[\sigma (\mathds{1}\otimes\sigma_2^G)]\nonumber\\
&=& \lim\limits_{V\rightarrow\infty}\,\frac{V+1}{2}\,\textrm{Tr}[\sigma(\rho_{\mathrm{th}}^V\otimes\sigma_2^G)],
\label{trick}
\end{eqnarray}
\vspace{5pt}

\noindent where
$
\rho_{\mathrm{th}}^V=\frac{2}{V+1}\sum_n\left(\frac{V-1}{V+1}\right)^n\ket{n}\bra{n}
$
is a thermal state with a covariance matrix $V\mathds{1}$. Identity (\ref{trick}) holds because this state multiplied by $(V+1)/2$ tends to $\mathds{1}$ when $V$ tends to infinity. Next we express $\sigma$ as a result of the transformation of the initial state $\rho_{\mathrm{in}}\otimes|0\rangle\langle0|$ by the circuit in Fig.~\ref{Figure}:
\begin{eqnarray}
\textrm{Tr}&&\hspace{-10pt}[\sigma(\rho_{\mathrm{th}}^V\otimes\sigma_2^G)]\hspace{-18pt}\\
 &\hspace{-18pt}=&\hspace{-8pt} \textrm{Tr}\left[ U_S\,T[(\Phi_{\eta}\otimes\Phi_{\mu})[U_{\lambda}(\rho_{\mathrm{in}}\otimes\ket{0}\bra{0})U_{\lambda}^{\dag}]]U_S^{\dag}\,\,\rho_{\mathrm{th}}^V\otimes\sigma_2^G \right]\nonumber\\
&\hspace{-18pt}=& \hspace{-8pt}\textrm{Tr}\left[ (\rho_{\mathrm{in}}\otimes\ket{0}\bra{0})U_{\lambda}^{\dag}(\Phi_{\eta}\otimes\Phi_{\mu})[T[U_S^{\dag}(\rho_{\mathrm{th}}^V\otimes\sigma_2^G)  U_S]]U_{\lambda}\right].\nonumber
\label{trths2}
\end{eqnarray}
Here $U_{S}$ is the final symplectic transformation, $U_{\lambda}$ describes the action of the two-mode squeezer,  $T$ is the partial transposition in the second mode, and $\Phi_{\eta}$ ($\Phi_{\mu}$) denotes the additive Gaussian noise channel with the noise variance $\eta$ ($\mu$) being applied to the first (second) mode. 
At the final step we  used the invariance of the trace under cyclic permutations and the equivalence  of the partial transposition and additive noise channel to their duals with respect to the scalar product of operators defined as $\langle\langle A|B\rangle\rangle = \tr[A^\dag B]$ on a set of density operators of two-mode states. Let us prove the last two statements.

\begin{itemize}
\item The partial transposition as a map defined on the set of density operators is equal to its dual.

{\it Proof:} Let us take a representation of two arbitrary density operators describing bipartite states in some basis $\rho=\sum_{ijkl}c_{ijkl}\ket{ij}\bra{kl}$ and $\sigma=\sum_{nmrs}d_{nmrs}\ket{nm}\bra{rs}$ and apply partial transposition $T$ on $\rho$. Then we have:

\begin{eqnarray}
\textrm{Tr}[\,T[\rho]\,\sigma]&=&\textrm{Tr}\left[\sum_{ijkl}c_{ijkl}\ket{il}\bra{kj}\sum_{nmrs}d_{nmrs}\ket{nm}\bra{rs}\right]\nonumber\\
&=&\sum_{ijkl}c_{ijkl}d_{kjil}\nonumber\\
&=&\textrm{Tr}\left[\sum_{ijkl}c_{ijkl}\ket{ij}\bra{kl}\sum_{nmrs}d_{nmrs}\ket{ns}\bra{rm}\right]\nonumber\\
&=&\textrm{Tr}[\rho\,T[\sigma]]. \quad \blacksquare
\end{eqnarray}

\item  The Gaussian additive noise (product) channel is equal to its dual on the set of density operators of two-mode states.

{\it Proof:} We prove first the equivalence on the example of the two-mode channel $\Phi_\eta\otimes\Phi_\mu$ applied to $\rho_{\mathrm{\mathrm{TMS}}}$:
 \begin{equation}
\tr\left[(\Phi_\eta\otimes\Phi_\mu)[\rho_{\mathrm{\mathrm{TMS}}}]\rho'\right]=\tr[\rho_{\mathrm{\mathrm{TMS}}}(\Phi_\eta\otimes\Phi_\mu)[\rho']].
\end{equation}
 Using Eq.~\eqref{rhonoise} and  the linearity of the trace we move it inside the integral and then make a cyclic permutation of the displacement operators. Then by applying the expression of the Hermitian conjugate of the displacement operator in the form $D^{\dag}(x,p)=D(-x,- p)$ and by changing  the variables $-x_1\rightarrow x_1$, $-p_1\rightarrow p_1$, $-x_2 \rightarrow x_2$, and $-p_2\rightarrow p_2$, we arrive at the desired conclusion.

This proof holds if we replace the state $\rho_{\mathrm{\mathrm{TMS}}}$ by an arbitrary density operator. 
$\blacksquare$
\end{itemize}

Following Eq.~\eqref{trths2} the trace on the right-hand side of Eq.~\eqref{trick} can be computed as the trace of the product of density matrices when the dual circuit is applied to the state $\rho_{\mathrm{th}}^V\otimes\sigma_2^G $ taking into account the properties of the dual maps discussed above. The big advantage of doing so is that $\rho_{\mathrm{th}}^V\otimes\sigma_2^G $ is a Gaussian state and all the transformations that constitute the circuit (and their duals) preserve the Gaussian character of the state. 
Thus the state $\rho^*=U_{\lambda}^{\dag}(\Phi_{\mu}\otimes\Phi_{\eta})[T[U_S^{\dag}\rho_{\mathrm{th}}^V\otimes\sigma_2^G  U_S]]U_{\lambda}$ is completely determined by its covariance matrix $\gamma^*$,
which is the result of the application of the dual circuit to the covariance matrix of $\rho_{\mathrm{th}}^V\otimes\sigma_2^G $,
\begin{equation}
\resizebox{.99\hsize}{!}{$\gamma^*=S_{\mathrm{\mathrm{TMS}}}(-\lambda)\Bigg(T\Big[S^T\begin{pmatrix}V\mathds{1}&&0\\0&&\nu_-\mathds{1}\end{pmatrix}S\Big]+\begin{pmatrix}\eta\mathds{1}&&0\\0&&\mu\mathds{1}\end{pmatrix}\Bigg)S_{\mathrm{\mathrm{TMS}}}^T(-\lambda)$},
\end{equation}
where $S$ represents the symplectic diagonalization that gives the symplectic eigenvalues, $T$ is the partial transposition (which acts on the  two-mode covariance matrices as $p_2\rightarrow-p_2$) and $S_{\mathrm{\mathrm{TMS}}}(\lambda)=\frac{1}{\sqrt{1-\lambda^2}}\begin{pmatrix}\mathds{1}&&\lambda\sigma_z\\\lambda\sigma_z&&\mathds{1}\end{pmatrix}$,  the two-mode squeezing transformation.

From covariance matrix  $\gamma^*$, we can easily deduce the Wigner function $W^*(x_1,p_1,x_2,p_2)$ of $\rho^*$ and with its help compute the following trace, which is equal to the trace in Eq.~\eqref{trths2}:
\begin{eqnarray}
\textrm{Tr}[(\rho_{\mathrm{in}}&\otimes&\ket{0}\bra{0})\,\rho^*]\\
 &=& (2\pi)^2\int dx_1dp_1dx_2dp_2\,W_{\rho_{\mathrm{in}}}(x_1,p_1)\nonumber\\
&  &\times \,W_{\ket{0}}(x_2,p_2)W^*(x_1,p_1,x_2,p_2)\nonumber\\
& = & 2\pi\,C \,\int dx_1dp_1 \, W_{\rho_{\mathrm{in}}}(x_1,p_1)W_{\mathrm{th}}^m(x_1,p_1).\nonumber
\end{eqnarray}
Here the  normalization factor $C$ is obtained by integrating  over the  variables of the second mode (it is a simple Gaussian integral) :
\begin{equation}
2\pi\int dx_2dp_2W_{\ket{0}}(x_2,p_2)W^*(x_1,p_1,x_2,p_2)=C\,W_{\mathrm{th}}^m(x_1,p_1), 
\end{equation}
and by stressing out a new Wigner function $W_{\mathrm{th}}^m(x_1,p_1)~=~\frac{1}{ \pi m} e^{-(x_1^2+p_1^2)/m}$ corresponding to a thermal state of variance $m$. Returning to the state space, the computation of the trace can be carried out as follows:
\begin{eqnarray}
\textrm{Tr}[\rho_{\mathrm{in}}\otimes\ket{0}\bra{0}\,\rho^*]\
& =  & C \,\textrm{Tr}\Big[\sum_n\phi_n\ket{n}\bra{n}\rho_{\mathrm{th}}^m\Big]\\
 &= & C\,\frac{2}{m+1}\sum_{n=0}^{\infty}\phi_n\Big(\frac{m-1}{m+1}\Big)^{n}.\nonumber
\end{eqnarray}
  Both parameters, $m$ and $C$, depend on $V$. The explicit formulas for $m$ and $C$ are cumbersome and we do not present it here, however, they allow us  to carry out the limit $V\rightarrow\infty$ which provides the trace in Eq.~\eqref{trick}. Together with Eq.~\eqref{denom}, this gives us a value for $g_2$ following Eq.~\eqref{gauss}.




\begin{thebibliography}{99}
\bibitem{ad} Inequality (\ref{delta}) was derived  in \cite{duan} without using the PPT criterion. However, it is straightforward to derive the inequality by applying the PPT criterion on the parametric form of  the uncertainty relations introduced in \cite{duan} and
thus setting the correspondence with the results in \cite{simon}.
\end{thebibliography}
\end{document}